\documentclass[12pt,preprint]{aastex}

\usepackage{graphicx}
\usepackage{natbib}
\usepackage{pdflscape} %landscape
\usepackage{amssymb}
\usepackage{ulem} %to strike a line
\usepackage{color} %color text

\newcommand{\be}{\begin{equation}}
\newcommand{\ee}{\end{equation}}
\newcommand{\nn}{\mbox{} \nonumber \\ \mbox{} }
\newcommand{\ba}{\begin{eqnarray}}
\newcommand{\ea}{\end{eqnarray}}

\newcommand\eg{{\it{{e.g.,\ }}}}

\newcommand{\LC}{light cylinder}

\newcommand{\Bf}{{magnetic field}}

\newcommand{\ms}{magnetosphere}

\begin{document}

\title
{Relativistic Rotating Vector Model}

\author{Maxim Lyutikov}
\affil{Department of Physics, Purdue University, 
 525 Northwestern Avenue,
West Lafayette, IN
47907-2036, USA; lyutikov@purdue.edu
}

\date{Received/Accepted}
\begin{abstract}
The direction of polarization  produced by a moving source rotates with the respect to the rest frame. We show that this  effect, induced by pulsar rotation, leads to an important correction to polarization swings within the framework of  rotating vector model (RVM). 
We construct relativistic RVM taking into account finite heights of the emission region that lead to aberration, time-of-travel effects and relativistic rotation of polarization.   Polarizations swings at different frequencies can be used, within the assumption of the radius-to-frequency mapping,  to infer  emission radii and  geometry of  pulsars. 

\end{abstract}

\section{Classic Rotating Vector Model}
The Rotating Vector Model, RVM,  \cite{1969ApL.....3..225R} is a cornerstone of pulsar theory. It explains, at least qualitatively, polarization swings observed in many pulsars \citep[\eg][]{1975ApJ...196...83M}. But often observations show deviation from   model  \citep{2004A&A...421..681E}.  A number of attempts have been made to improve the model  either considering finite emission heights \citep[\eg]{1991ApJ...370..643B} or distortions of the \ms\ \citep[\eg][]{2001ApJ...546..382H,2012ApJ...755..137C}. Another important effect is the rotation of the polarization direction due to the relativistic motion of the emitter \citep{1972NPhS..240..161C,1973ApJ...183..977F,1979ApJ...232...34B,2003ApJ...597..998L,2005MNRAS.360..869L,2004A&A...426..985V}.

Mathematical relations are best solved in a  rest frame of the pulsar.
 In this frame the magnetic moment is directed along $z$ axis, so that a unit $\mu$-vector is 
${\bf \mu} = \{0,0,1\}$. For a unit radius vector $\hat{{\bf r}} = \{ \cos \phi \sin \theta, \sin \phi \sin \theta, \cos \theta \}$ the unit vector along the \Bf\ at a point $\{\theta, \phi \} $ is 
\be
{\bf b} = \left\{\frac{3 \sqrt{2} \sin \theta  \cos \theta  \cos \phi }{\sqrt{3 \cos (2
   \theta )+5}},\frac{3 \sqrt{2} \sin \theta  \cos \theta  \sin \phi }{\sqrt{3 \cos
   2 \theta +5}},\frac{3 \cos 2 \theta +1}{\sqrt{6 \cos 2 \theta +10}}\right\}
   \label{bb}
   \ee
   Introducing rotation $L_t$ and inclination $L_\alpha$ operators,
  \be
  L_t = \left(
\begin{array}{ccc}
 \cos \Omega t  & -\sin \Omega t  & 0 \\
 \sin \Omega t  & \cos \Omega t  & 0 \\
 0 & 0 & 1 \\
\end{array}
\right)
, \, 
  L_\alpha =    \left(
\begin{array}{ccc}
 \cos \alpha  & 0 & -\sin \alpha  \\
 0 & 1 & 0 \\
 \sin \alpha  & 0 & \cos \alpha  \\
\end{array}
\right),
\ee
the line of sight in that frame is 
  \be
  {\bf n} = L_\alpha \cdot L_t \cdot {\bf n}_0
 , \, 
  {\bf n}_0 = \{ \sin\theta_{ob} ,0, \cos \theta_{ob} \}
  \label{1}
  \ee
  while the plane of the sky is determined by two vectors
  \ba && 
  {\bf l} = L_t \cdot L_\alpha \cdot  {\bf l}_0
  \nn &&
  {\bf m} = L_t \cdot L_\alpha  \cdot {\bf m}_0
  \nn &&
  {\bf l}_0 = \{ 0,1,0\}
   \nn &&
    {\bf m}_0 =  \{ \cos \theta_{ob},0,-\sin \theta_{ob} \}
    \ea
    In the above relations $\alpha$ is the pulsar inclination angle (angle between rotation and magnetic axes), $\theta_{ob}$ is the viewing angle -  (angle between rotation axis and the line of sight).
    
RVM  assumes that emission is directed align the local \Bf, thus the condition ${\bf b} \parallel {\bf n}$ determines the coordinates $\{\theta,\, \phi\}$ of a  point  in the \ms\ that contributes to the emission. Conventionally (assuming that emission occurs at $r=0$)  the requirement ${\bf b} \parallel {\bf n_0}$ gives
 \ba &&
 \tan \phi =\frac{\sin\theta _{{ob}} \sin \Omega t }{\cos \alpha 
   \sin\theta _{{ob}} \cos \Omega t -\sin \alpha  \cos
  \theta _{{ob}}}
  \nn &&
  \frac{3 \cos 2 \theta +1}{\sqrt{6 \cos 2 \theta +10}}=\cos  \alpha  \cos
    \theta _{{ob}}+\sin  \alpha  \sin  \theta _{{ob}}
   \cos \Omega t
   \label{phi}
   \ea
   For given parameters of the pulsar and the LoS these two equations determine the emission point  $\{\theta,\, \phi\}$ that contributes to the emission.
   
  To find the polarization direction we note that within the RVM the polarization is perpendicular to the azimuthal vector 
   \be
   {\bf e}_ \phi = \{ - \sin \phi,\cos \phi,0\}
   \label{ephi}
   \ee
and to the LoS ${\bf n}$.  Thus polarization is along
   \be
   {\bf e} _p = {\bf e}_ \phi  \times{\bf n}
   \ee
   ($ {\bf e} _p$ can be normalized, but the normalization factor cancels out if we calculate $\tan \chi$.)
The polarization angle is
   \be
   \tan  \chi = \frac{{\bf e} _p \cdot {\bf l}  }{{\bf e} _p \cdot {\bf m}} 
   \label{tanchi}
   \ee
   evaluates to
   \be
 \tan  \chi    =
   -\frac{\sin \alpha  \sin \phi  \sin\theta _{{ob}}+\cos \alpha 
   \sin \phi  \cos\theta _{{ob}} \cos \Omega t -\cos \phi  \cos
  \theta _{{ob}} \sin \Omega t }{\cos \alpha  \sin \phi  \sin 
   \Omega t +\cos \phi  \cos \Omega t }
   \ee
   Using Eq. (\ref{phi}) to eliminate $\phi$ we find \citep{1969ApL.....3..225R}
   \be
   \tan (\chi )=\frac{\sin \alpha  \sin \Omega t }{\sin \alpha  \cos \theta
   _{{ob}} \cos \Omega t -\cos \alpha  \sin \theta
   _{{ob}}}
   \label{RVM}
   \ee
   
   \section{Relativistic effects}
   
      \subsection{Aberration}

   Finite distance of the emission region from the center of the star introduces a number of corrections to the classic RVM. Some of these corrections - aberration - have been considered previously \citep[\eg][]{1991ApJ...370..643B}. Here we include another important relativistic effect: rotation of polarization produced by a moving source. (Below the radius $r$ is normalized to the light cylinder radius $r$; we use spherical system of coordinates, so $r=1$ does not mean a point is located on the \LC; speed of light is set to unity).
   
   In the chosen frame the angular and the linear  velocities and  are
   \ba &&
  {\bf  \Omega} =  L_t \cdot L_\alpha \cdot  {\bf \Omega}_0
   \nn &&
   {\bf \Omega}_0=\{0,0,\Omega\}
   \nn &&
   v= r  {\bf  \Omega} \times \hat{{\bf r}} 
   \ea
   
\begin{figure}[h!]
 \centering
 \includegraphics[width=.32\columnwidth]{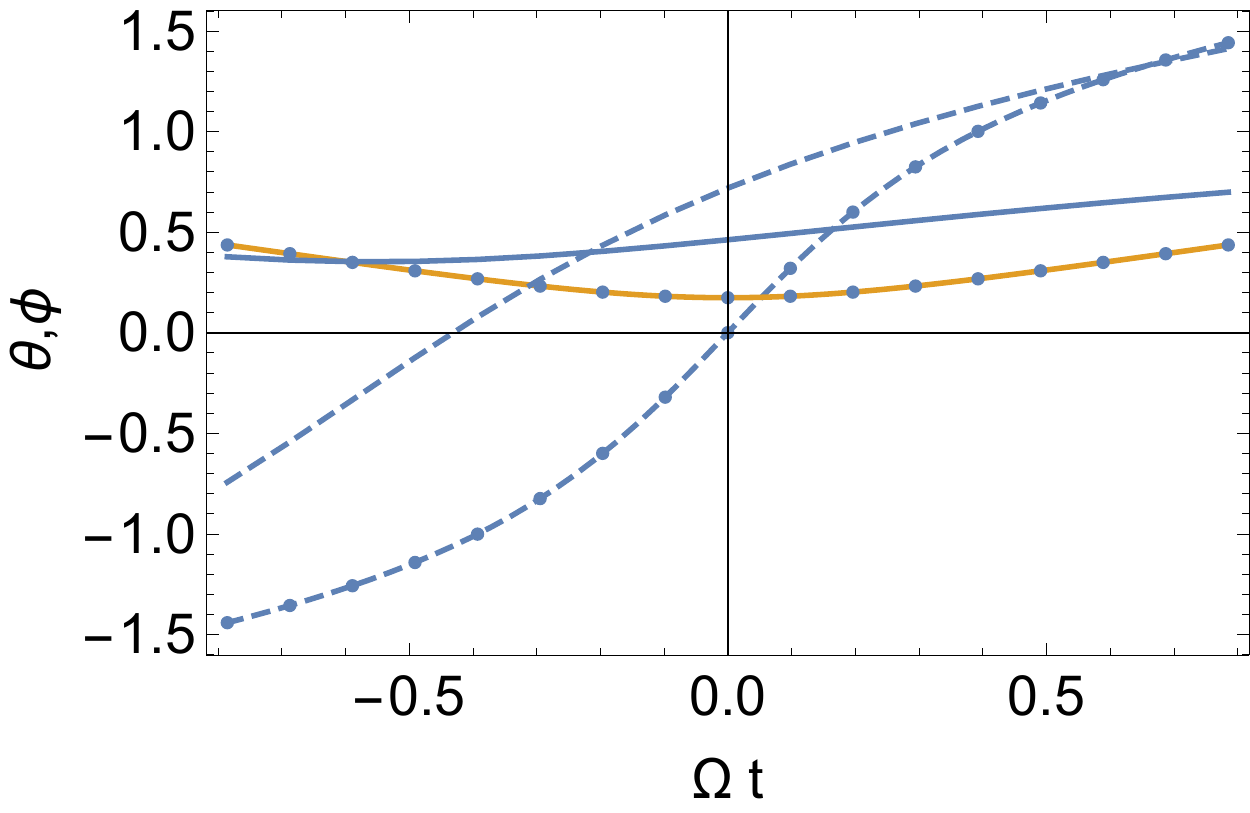}
 \includegraphics[width=.32\columnwidth]{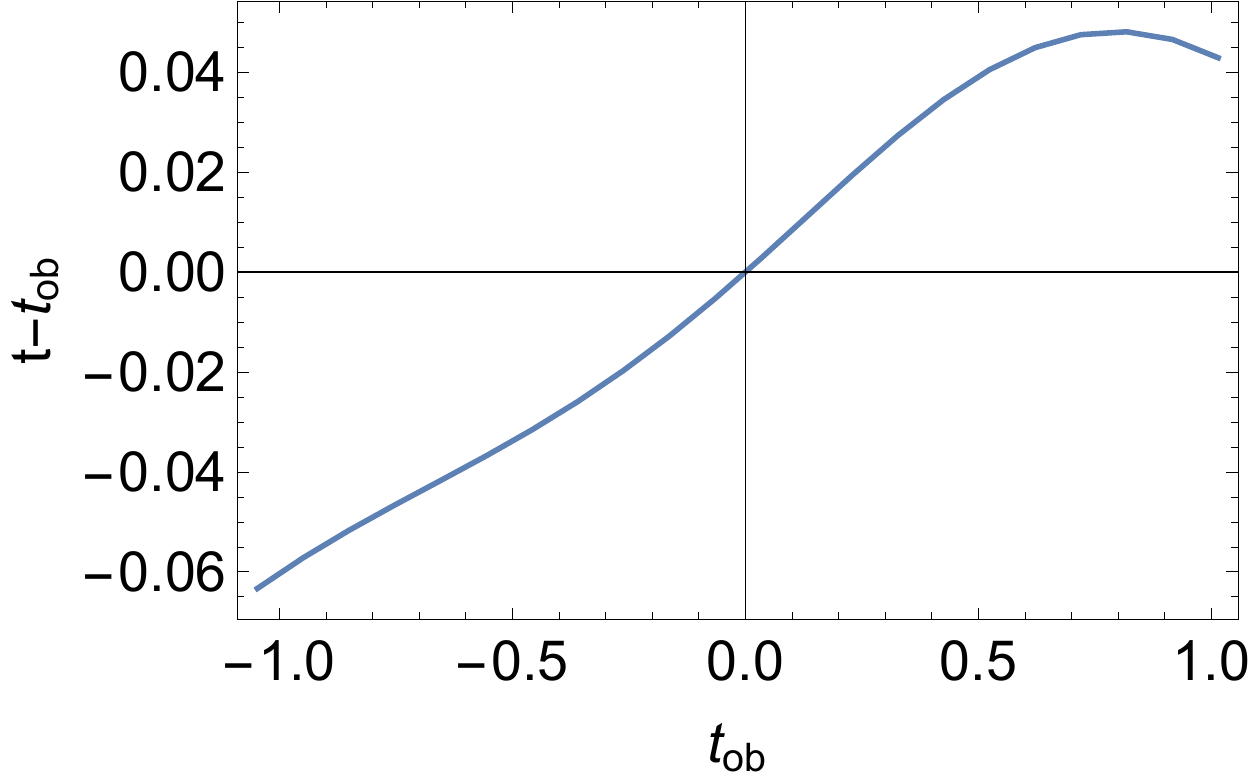}
  \includegraphics[width=.32\columnwidth]{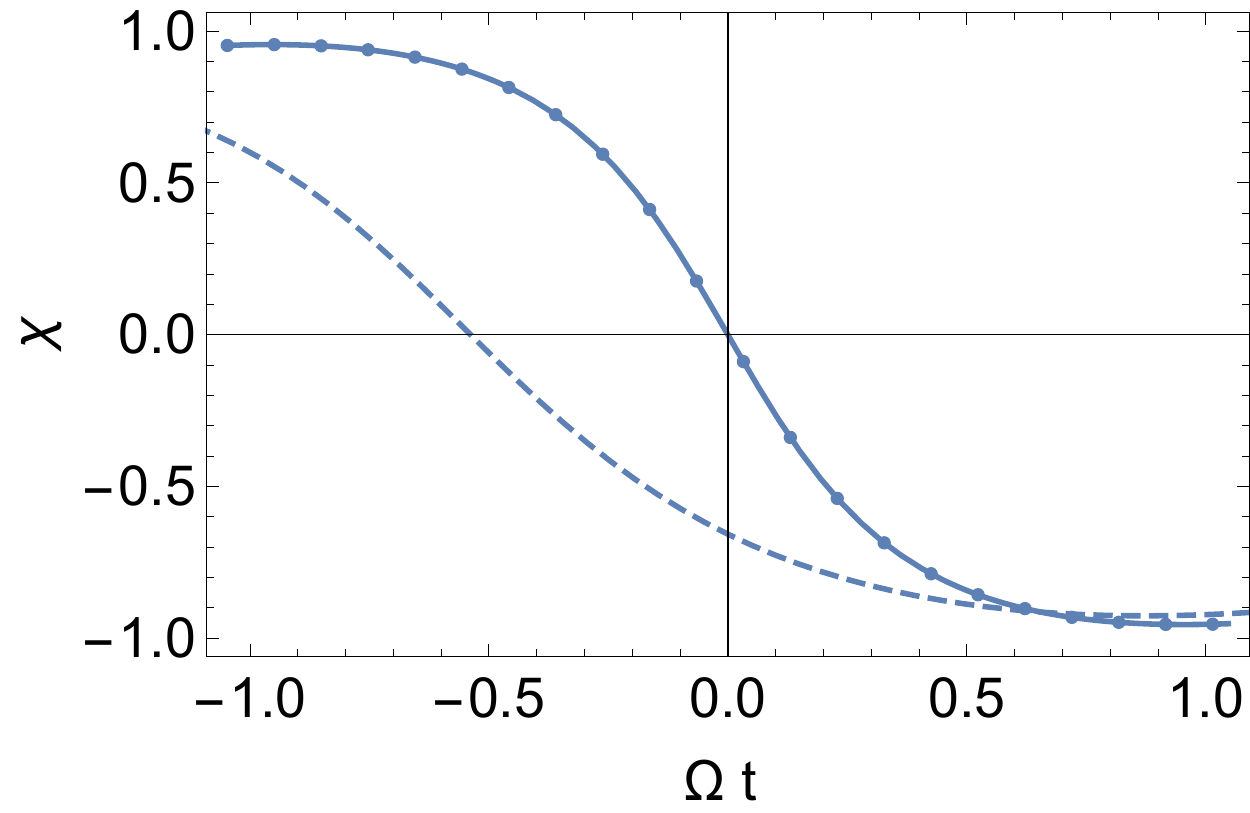}
 \caption{Various relativistic corrections for the case of inclination angle $\alpha = \pi/4$, viewing angle $\theta_{ob}=\pi/3$ distance to the star $r=1$ . {\it Left Panel}: Effects of aberration. 
 Plotted are the location of the points in the \ms\ contributing to emission.
 Solid lines are $\theta(t)$, dashed lines are $\phi (t)$. Lines with dots are the conventional RVM, dots are numerical results (agreement of numerical results with the analytic RVM also serves as a test of our numerical scheme).
 {\it Center Panel}: time-of-travel effects. Plotted is the difference between the coordinate and the observer time versus the observer time. We see that time-of-flight corrections are typically small for the parameters chosen. 
 {\it Right Panel}: Classic RVM (solid line with dots corresponding to the numerical solutions of Eq.  (\protect\ref{RVM})) and the  effects of relativistic rotation of polarization for emission heigh $r=1$ (dashed line).}
 \label{angle-abb}
\end{figure}

   The first relativistic effects is aberration: in the pulsar frame the wave is emitted along
\be
{\bf n}_1 = \frac{{\bf n}- {\bf v} \gamma \left(1- \frac{\gamma}{1+\gamma} (\bf{n} \cdot \bf{v}) \right)}{\gamma (1-(\bf{n} \cdot \bf{v})) }
\label{n1}
\ee
The condition ${\bf n}_1 \parallel {\bf b}$ then determines $\theta$ and $\phi$ of the emission point. This condition is difficult to resolve analytically even for small $r$, so we resort to numerical methods, Fig. \ref{angle-abb}

  \subsection{Time-of-flight effects}

Finite height of emission will also affect polarization sweeps due to time of travel effects. 
%The observer time is related to the coordinate time as
%\be
%\frac{\partial t_{ob}} {\partial t}= 1 - \beta \cos \theta_{v-LoS}
%\ee
%where $\theta_{v-LoS}$ is the angle between the direction of motion and the LoS
Let us calculate it with  respect to the plane parallel to the plane of the sky and passing through the pulsar. The corresponding observer time is  
$
t_{ob} = t - r (\hat{{\bf r}}  \cdot {\bf n})
$. For convenience we shift the observer times $t_{ob}$  by $t_0$ so that $t=0$ corresponds to $t_{ob}=0$,
\be
t_{ob} = t - r (\hat{{\bf r}}  \cdot {\bf n})+t_0
\label{tob}
\ee
For a given observer time $t_{ob}$ we invert relation (\ref{tob}) to find the emission time $t$, see Fig. \ref{angle-abb}, Center Panel.

\subsection{Rotation of polarization from moving source}

Finally, the most important  effect that has previously been missed is  rotation polarization direction due to the motion of the source. This effect has previously been considered for the case of  synchrotron emission by relativistically moving sources in AGNe \cite{2005MNRAS.360..869L} and GRBs \cite{2003ApJ...597..998L}. We can use the result of \cite{2003ApJ...597..998L} with the following substitution. 
In case of synchrotron emission the direction of polarization {\it in the plasma rest frame} is orthogonal both to the direction of wave propagation and the projection of the \Bf\ onto the plane orthogonal to the direction of propagation. In case of curvature emission (which is assumed as a basic emission mechanism for the rotating vector model) the direction of polarization in the pulsar frame is orthogonal to the photon propagation and {\it the vector normal to the plane of the \Bf} - for purely dipole field this is the azimuthal vector ${\bf e}_\phi $, Eq. (\ref{ephi}).

Thus, we can use the relativistic polarization transformation, Eq. (3) of \cite{2003ApJ...597..998L}, substituting ${\bf B}' \rightarrow {\bf e}_ \phi$; we find
\ba &&
{ {\bf e}_p} ={  {\bf n} \times {\bf q}' \over 
\sqrt{ q^{\prime 2} - ( {\bf n} \cdot {\bf q}')^2} } \mbox{,}
\nn &&
{\bf q}' = {\bf e}_\phi +
 {\bf n}  \times (  {\bf v} \times  {\bf e}_\phi) 
-{ \gamma \over 1+\gamma} ( {\bf e}_\phi  \cdot {\bf v} ) {\bf v} \mbox{.}
\label{ee0}
\ea
Equation (\ref{ee0}) also demonstrates that  for the purely poloidal motion ($ {\bf e}_\phi  \cdot {\bf v} =0$) along the emission direction does not affect the polarization - for ${\bf v} \parallel {\bf n}$ we have ${\bf q}' = {\bf e}_\phi $. 

We have formally solved the relativistic polarization in the framework of the RVM: for a given direction to the observer frame, Eq. (\ref{1}), Eq. (\ref{n1}) gives the direction in the rotating pulsar frame; it must be aligned with the local \Bf\, Eq. (\ref{bb}). This condition gives two angular coordinates $\{\theta, \phi\}$ corresponding to the emission point. Next, Eq. (\ref{ee0}) gives the direction of the polarization vector and Eq. (\ref{tanchi}) gives the polarization angle.

\subsection{Putting it all together}

\begin{figure}[h!]
 \centering
 \includegraphics[width=.9\columnwidth]{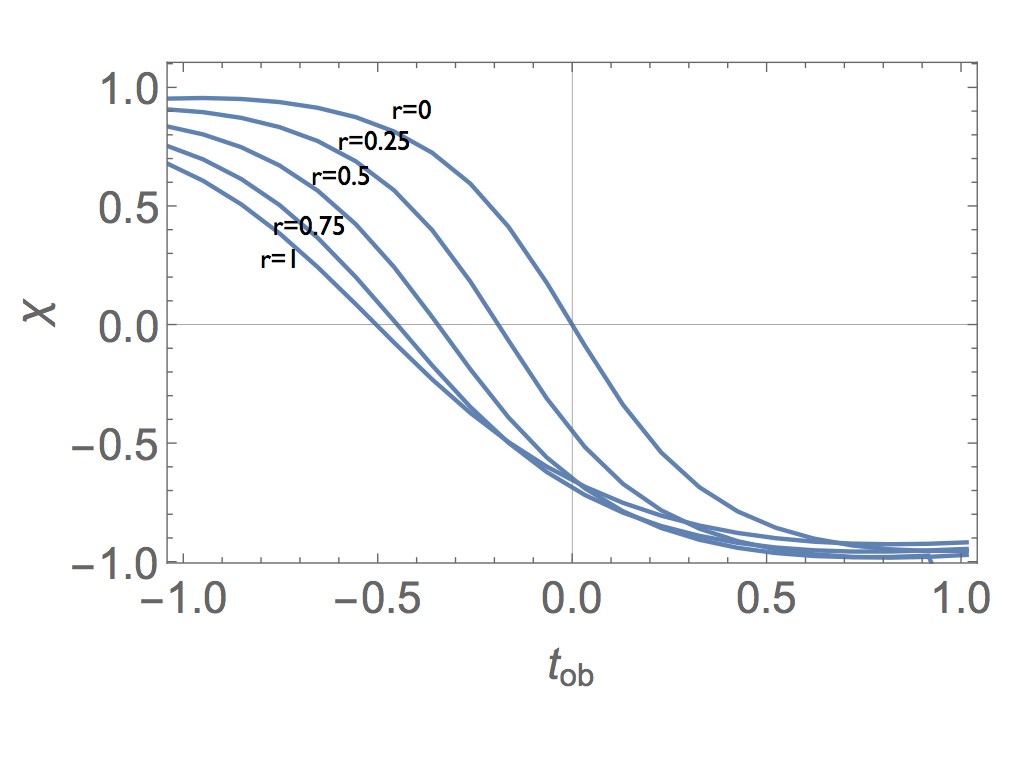}
  \caption{Polarization angle sweeps for different emission radii taking into account all the relativistic effects: aberration, time of flight and rotation of polarization. Distance is measured in the \LC\ radii.  Inclination angle $\alpha = \pi/4$, viewing angle $\theta_{ob}=\pi/3$, $r=1$.
 }
 \label{chiofrandtob}
\end{figure}

Next we put all the relativistic effects together: aberration, time of flight, and rotation of polarization, Fig. \ref{chiofrandtob}.

\begin{figure}[h!]
 \centering
 \includegraphics[width=.49\columnwidth]{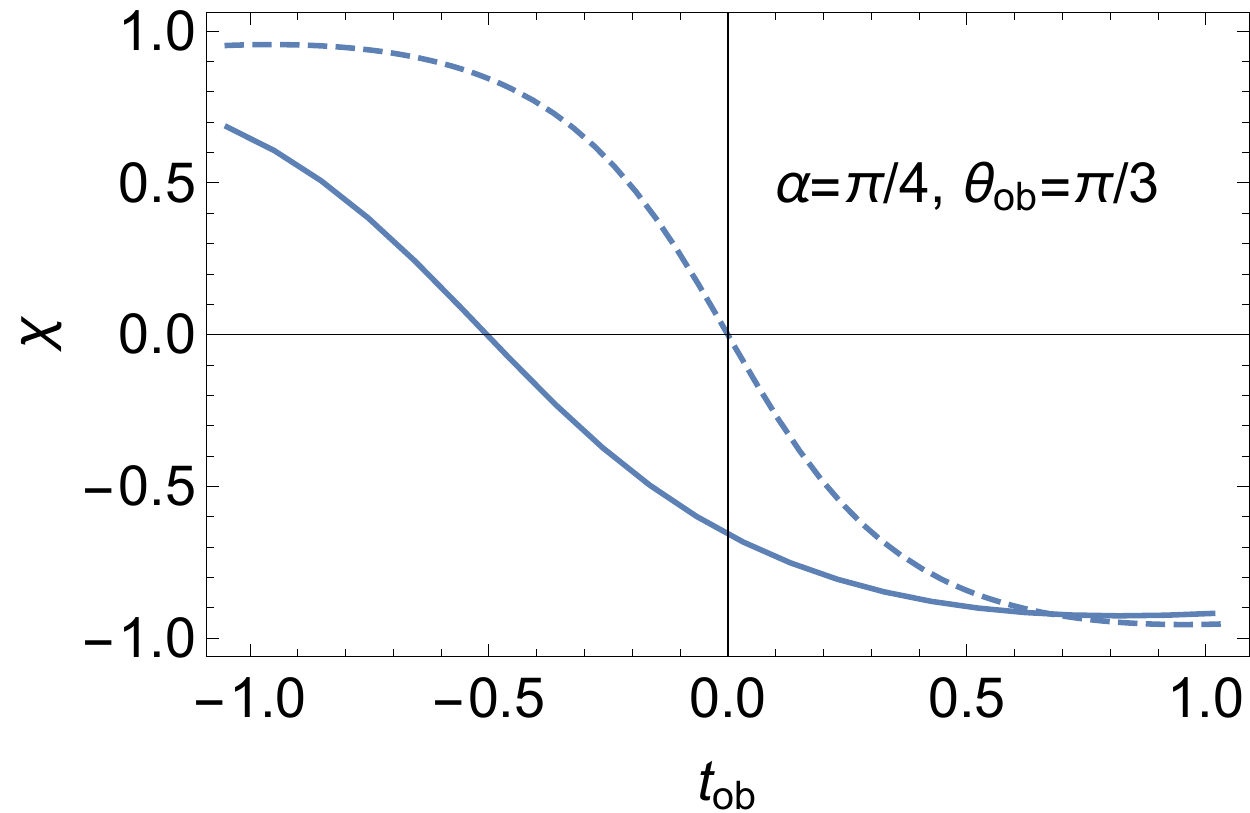}
 \includegraphics[width=.49\columnwidth]{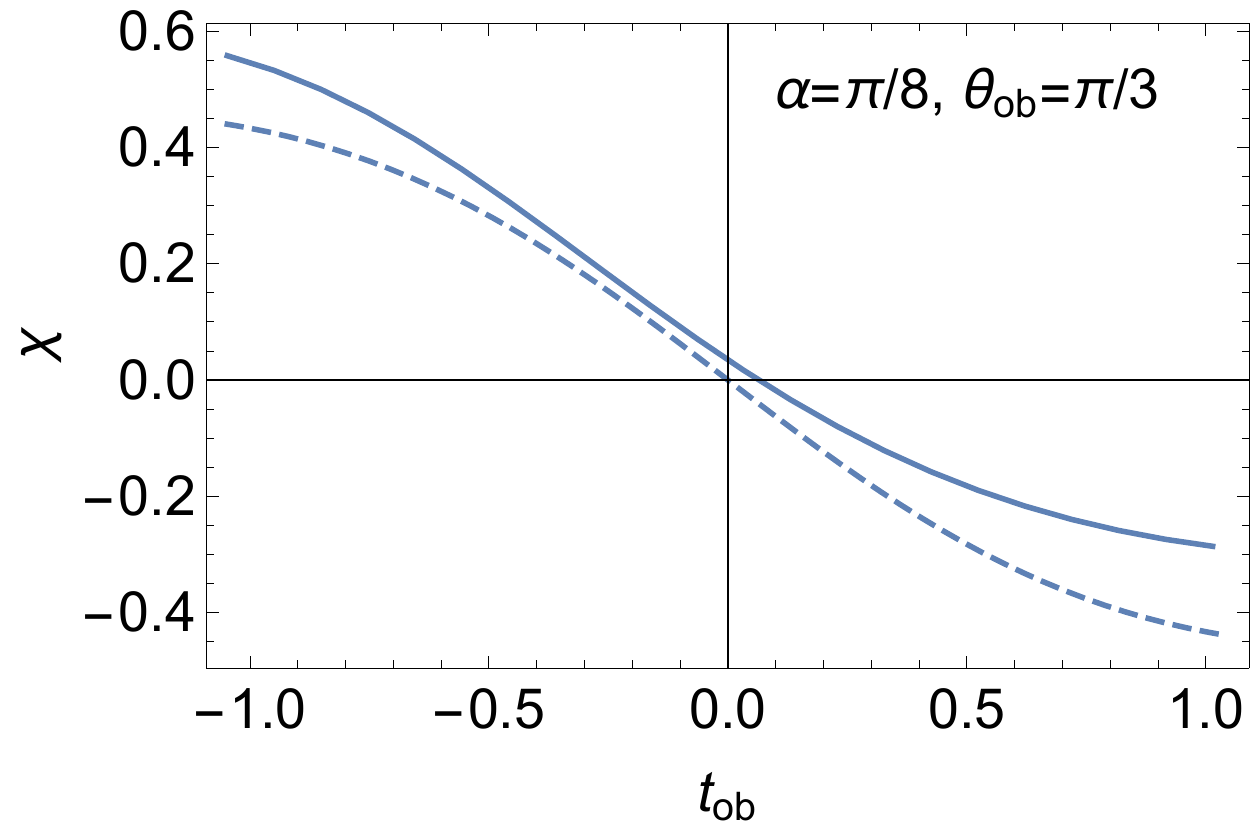}
\\
\includegraphics[width=.49\columnwidth]{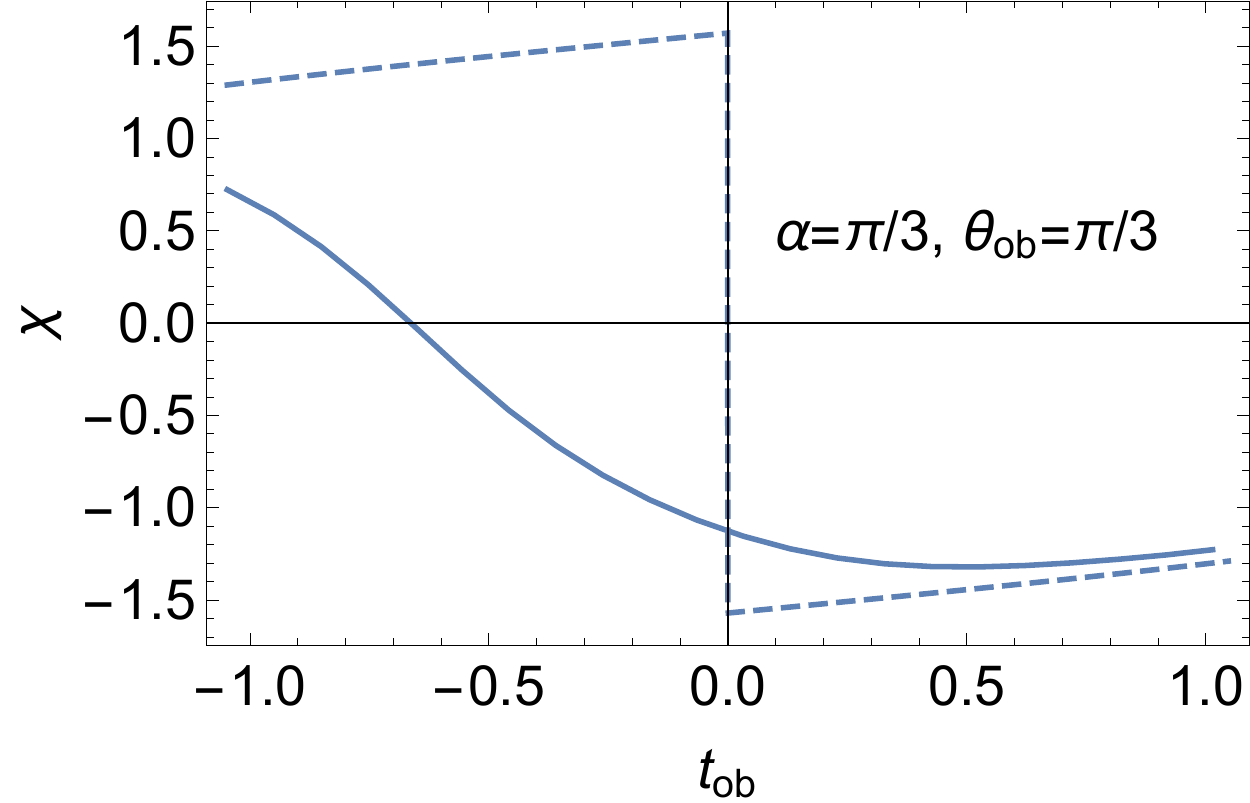}
\includegraphics[width=.49\columnwidth]{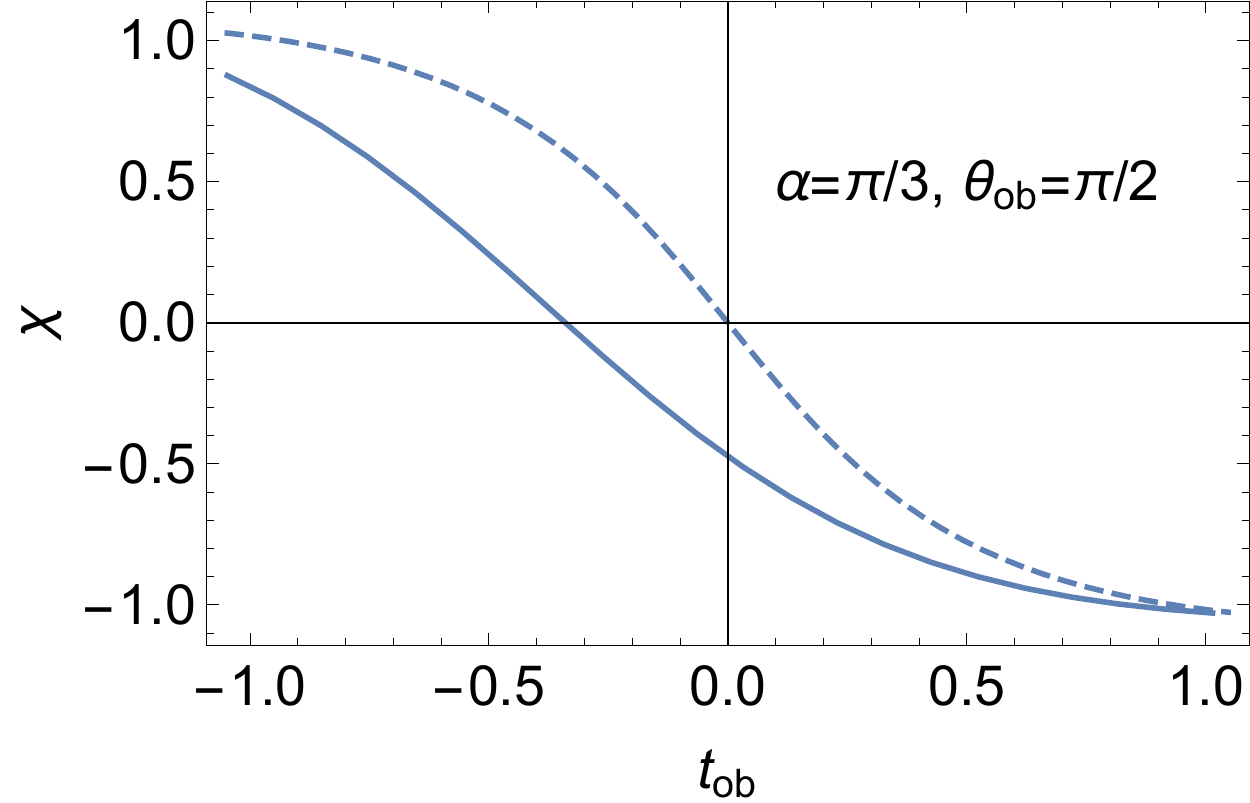}

  \caption{Polarization angle sweeps for different parameters; $r=1$.
 }
 \label{chi-swings}
\end{figure}

A few examples of polarization swings for various parameters are given in Fig. \ref{chi-swings}

\subsection{A simple example}

The above relations are highly complicated and cannot be resolved  analytically even in the limit of small emission radii. Let us illustrate the principle using an over-simplified  example: an aligned rotator, $\alpha =0$. The conventional RVM then simply gives, Eq. (\ref{phi}), $\chi=0$ - polarization is aligned with the projection of the rotation axis onto the plane of the sky. Relations (\ref{phi}) give in this case 
\ba &&
\tan \theta_{ob} = \frac{3 \sin 2 \theta }{3 \cos 2 \theta +1}
\nn &&
 \phi = \Omega t
 \label{22}
 \ea
 
 For finite emission radii, taking into account aberration (in the limit $r \rightarrow 0$) gives
 \be
  \phi = \Omega t - r \Omega \frac{\sin \theta} {\sin \theta_{ob}}
  \ee
 while relation between $\theta $ and $\theta_{ob}$ remains the same as (\ref{22}).
 
The polarization angle becomes
 \be
 \tan \chi \approx r \Omega \frac{\sin \theta} {\sin \theta_{ob}} = r \Omega  \left(\cos \theta -\frac{\sec \theta }{3}\right) \approx\frac{ r \Omega}{6} , \mbox{for }  \theta_{ob} \ll 1
  \ee
% Since  $\sin \left(\theta _{\text{ob}}\right)=\frac{3 \sqrt{2} \sin \theta  \cos \theta }{\sqrt{3 \cos 2 \theta +5}}$ this becomes
  
\section{Discussion}

The key point of the present paper is to point out that an important relativistic effects --  rotation of the  direction of polarization emitted by a moving source  -- has been previously missed in the modifications of RVM.  The effect is linear in $r \Omega$, as well as aberrations and time of travel effects. As Fig. \ref{chiofrandtob} demonstrates, the relativistic effects become important, somewhat unexpectedly, at fairly small radii, $25\%-50\%$ of the \LC. These effects are bound to be dominant in millisecond pulsars. We foresee that these effects will also be   important in modeling the light curves (and radio polarization swings) in the $\gamma$-ray pulsars \citep{2013ApJS..208...17A}.

Fig. \ref{chiofrandtob} also demonstrates that polarizations swings measured at different frequencies can constrain the radius-to-frequency mapping. Or inversely,  polarizations swings at different frequencies can be used to infer the emission radius and the geometry of a particular pulsar.   One of the major constrains is the assumption of the dipolar structure Ð at larger radii the sweep-back of the fields lines and the distortion of the poloidal structure would become important. 
 We leave these applications to a future work. 
 
 I would like to thank Carl Gwinn and Yuri  Poutanen for discussions  and Universitat Autonoma de Barcelona  for hospitality.

\bibliographystyle{apj}
%\bibliography{liter}
% %\begin{comment}
%%\bibliographystyle{apj}
\bibliography{/Users/maxim/Home/Research/BibTex}

   \end{document}